\documentclass[sigconf,nonacm]{acmart}
\AtBeginDocument{%
  }

\usepackage{xspace}

\makeatletter
\DeclareRobustCommand\onedot{\futurelet\@let@token\@onedot}
\def\@onedot{\ifx\@let@token.\else.\null\fi\xspace}

\def\eg{\emph{e.g}\onedot} \def\Eg{\emph{E.g}\onedot}
\def\ie{\emph{i.e}\onedot}

\def\etal{\emph{et al}\onedot}
\makeatother

\definecolor{redcolor}{RGB}{215,25,28}
\definecolor{violetcolor}{RGB}{239, 66, 245}

\newcommand{\ourmethodspace}{SEQ+MD }
\begin{document}

\title{SEQ+MD: Learning Multi-Task as a SEQuence with Multi- Distribution Data}
\author{Siqi Wang}
\authornote{Work done during an internship at Etsy.}
\affiliation{%
  \institution{Boston University}
  \city{Boston}
  \country{USA}}
\email{siqiwang@bu.edu}

\author{Audrey Zhijiao Chen}
\affiliation{%
  \institution{Etsy}
  \city{Brooklyn}
  \country{USA}}
\email{achen@etsy.com}

\author{Austin Clapp}
\affiliation{%
  \institution{Etsy}
  \city{Brooklyn}
  \country{USA}}
\email{aclapp@etsy.com}

\author{Sheng-Min Shih}
\affiliation{%
  \institution{Etsy}
  \city{Brooklyn}
  \country{USA}}
\email{sshih@etsy.com}

\author{Xiaoting Zhao}
\affiliation{%
  \institution{Etsy}
  \city{Brooklyn}
  \country{USA}}
\email{xzhao@etsy.com}

\begin{abstract}
In e-commerce, the order in which search results are displayed when a customer tries to find relevant listings can significantly impact their shopping experience and search efficiency. Tailored re-ranking system based on relevance and engagement signals in E-commerce has often shown improvement on sales and gross merchandise value (GMV).
Designing algorithms for this purpose is even more challenging when the shops are not restricted to domestic buyers, but
can sale globally to international buyers. Our solution needs to incorporate shopping preference and cultural traditions in different buyer markets.
We propose the SEQ+MD framework, which integrates sequential learning for multi-task learning (MTL) and feature-generated region-mask for multi-distribution input. This approach leverages the sequential order within tasks and accounts for regional heterogeneity, enhancing performance on multi-source data.
Evaluations on in-house data showed a strong increase on the high-value engagement including add-to-cart and purchase while keeping click performance neutral compared to state-of-the-art baseline models.
Additionally, our multi-regional learning module is "plug-and-play" and can be easily adapted to enhance other MTL applications.
\end{abstract}



\keywords{Multi-task Learning, Mixed-distribution Learning, E-commerce Search, E-commerce Ranking}


\maketitle

\section{Introduction}
\begin{figure}[t]
\centering
\includegraphics[width=\columnwidth]{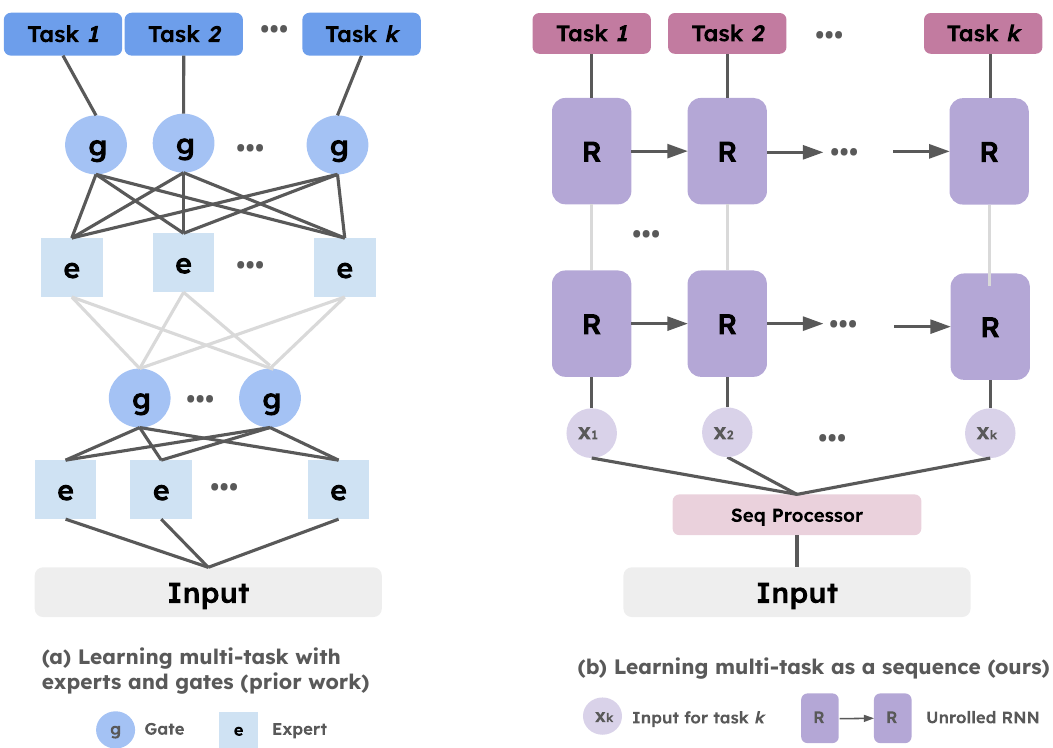}
\caption{
\textbf{MTL Architecture Comparison.} (a) Prior work~\cite{mlmmoe_ma2018modeling, ple_tang2020progressive, li2023adatt} uses \textit{experts} and \textit{gates} for task knowledge sharing, with variations in whether the expert or gate is shared among tasks. (b) Our SEQ learns multi-task as a sequence, where task knowledge is shared through sequence tokens.
}
\label{fig2_mtl_compare}
\end{figure}
In e-commerce, the design of listing display algorithms is crucial for enhancing the customer shopping experience. An effective search algorithm can significantly boost user engagement and drive increased revenue for the company~\cite{lari2022artifical}.
When a customer enters a query in the search window, the query typically goes through two stages to render final search results: retrieval and re-ranking. In the first stage, retrieval systems extract thousands of most relevant items from millions of listings; in re-ranking step, the thousands of listings are further re-ranked such that most relevant results are shown at the top.
Unlike traditional pattern-searching methods~\cite{sarwar2000analysis}, machine learning offers possibilities for more personalized search experiences~\cite{de2022machine, yoganarasimhan2020search}. 
The same search query from different users may yield completely different listing displays. 

However, designing such machine learning algorithms is challenging and involves two primary hurdles.
First, models often need to be simultaneously optimized on multiple tasks and objectives. 
For example, in e-commerce, customer engagement through clicks, or "window shopping," can inspire shopping ideas and lead
to purchases~\cite{oh2018clicking}. Therefore, predicting not only the listings with the highest purchase probability but also those with a high click probability can enhance user engagement.
Compared to training each task with a separate model, multi-task learning can enhance data utility and improve performance by sharing information across tasks~\cite{li2017better}. However, there is still room for improvement in training multiple tasks in a balanced manner and increasing effective communication between tasks~\cite{zhang2018overview, zhang2021survey}.

\begin{figure}[t]
\centering
\includegraphics[width=\columnwidth]{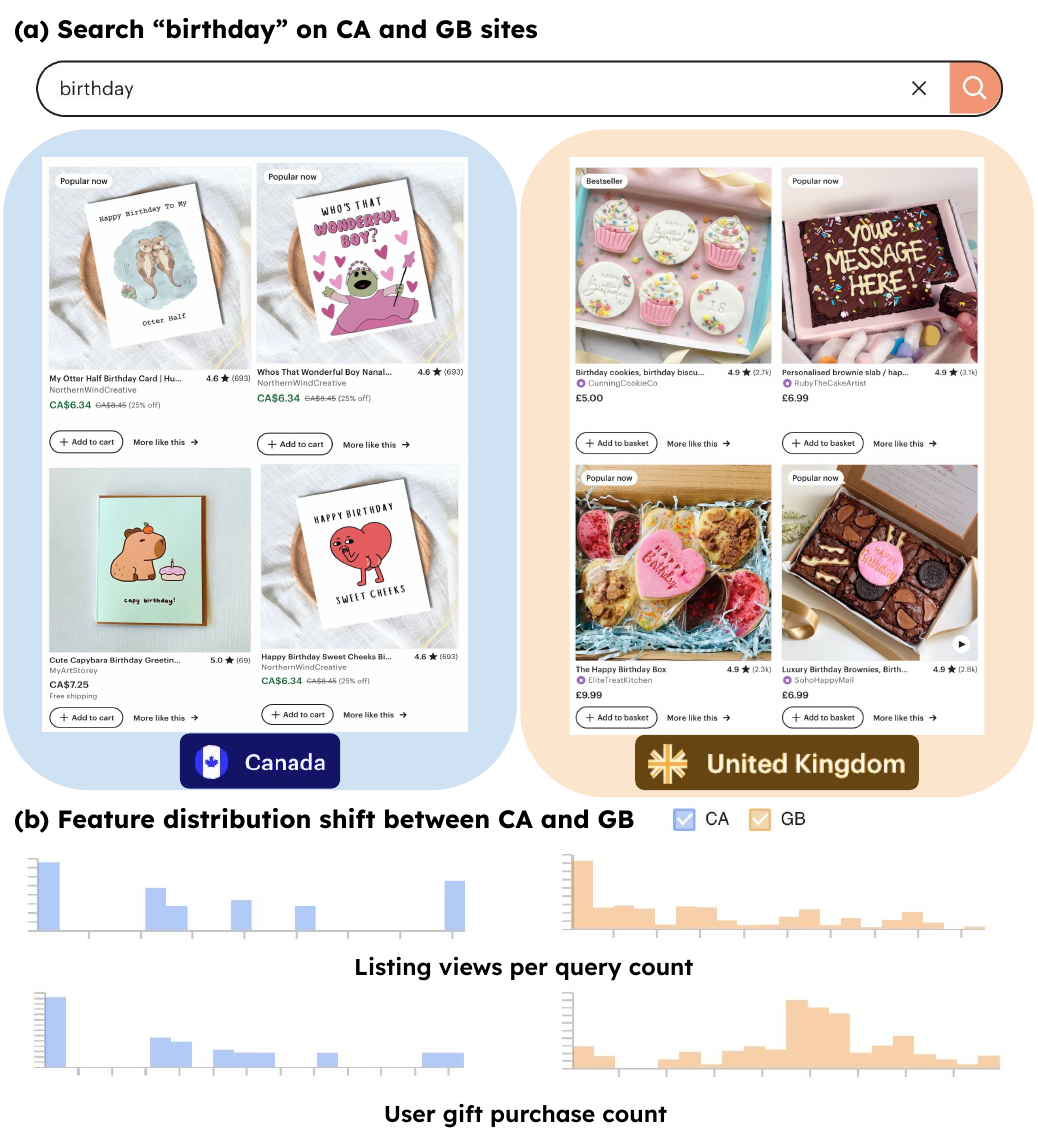}
\caption{
\textbf{Regional Difference Examples.} (a) The same search query on different regional sites should display different listings to reflect local preferences. For example, GB (United Kingdom) shoppers often choose cookie boxes as birthday gifts, while Canadian shoppers favor birthday cards.  (b) Feature distribution shifts across countries. In Canada (CA) and the UK (GB), some features display an entirely different distribution pattern, posing a challenge for the model to learn.
}
\label{fig1_motivation}
\end{figure}

The second challenge is that training data often contains regional variations, resulting in multiple distributions. This issue is particularly prominent in a global e-commerce marketplace, where shoppers have access to international inventory, not just local listings. Beyond the multilingual semantic matching problem (e.g., a French buyer searching for \textit{bijoux} should be able to find \textit{jewelry} listings from UK sellers), 
a key obstacle is addressing country-specific shopping preferences and cultural influences, which play a significant role in modeling consumer behavior.
For example, UK buyers purchase more cookie boxes as birthday gifts than buyers in other regions, as shown in Fig.~\ref{fig1_motivation}-(a). These regional differences can lead to distinct feature distributions and feature importance. As shown in Fig.~\ref{fig1_motivation}-(b), features such as the number of views for each listing per query vary by region. Some features might be uniformly distributed in certain regions (providing no helpful information for model prediction) but are vital in others. As a result, the expected search results can vary across countries. In the paper, we use the terms "country" and "region" interchangeably. Although we primarily focus on countries, the concept of a region can refer to any geographical size.

Existing methods usually address these two challenges separately. To the best of our knowledge, no single model currently solves both challenges effectively. Regarding multi-task learning (MTL), current methods treat tasks independently~\cite{shared_bottom_caruana1997multitask,misra2016cross, ruder2019latent}, ignoring their inherent sequential nature. Interaction between tasks, beyond a shared base model, is limited to shared experts or gates~\cite{li2023adatt, ple_tang2020progressive, mlmmoe_ma2018modeling}. For multi-region data, the current approach involves training a unified model irrespective of the region. However, as illustrated in Fig. 1, there are significant regional differences in raw input features. Incorporating regional factors could potentially enhance model performance, but training separate models for each region is sub-optimal due to an imbalanced data distribution, especially for regions with limited training data.

To this end, we propose the learning multi-task as a \textbf{SEQ}uence + \textbf{M}ulti-\textbf{D}istribution (\textbf{SEQ+MD}) framework , which can tackle the two challenges simultaneously. For the multi-task learning component, we recognize that tasks focusing on sequential actions can be naturally transformed into sequential learning problems. For example, predicting the probability of a "click" followed by a "purchase" makes intuitive sense, as users rarely purchase without first clicking on the listing. Thus, we propose learning multi-task sequences within our \textbf{SEQ} architecture as shown in Fig.~\ref{fig2_mtl_compare}-(b) 
For handling mixed input distributions, we separate input features into region-invariant and region-dependent groups. The region-dependent features are processed with a country embedding in our multi-distribution (\textbf{MD}) learning module, meaning these features are transformed according to their region, and then concatenated with the region-invariant features. An advantage of this approach is that the MD module is "plug-and-play" and can enhance the performance of any multi-task learning model on multi-source data.

We evaluated our framework on our in-house data offline and observed a 1.8\% performance increase in the critical purchase task while keeping the click task performance positive compared to baseline models. In summary, our contributions are:


\begin{itemize}
  \item  We introduced a new framework \textbf{SEQ} for multi-task learning leverage an improvised RNN architecture, specifically designed to handle tasks with sequential order, with a particular emphasis on those tasks with sparse data. SEQ not only extracts and utilizes the sequence relation between tasks, reduces redundant computations among related tasks but also demonstrates excellent transferability when adding new tasks. By decomposing a complex task into simpler, sequential tasks, SEQ effectively enhances the multi-task learning process.
  \item We developed a "plug-and-play" module \textbf{MD} for learning regional data with different distributions. The MD module enables the model to capture region-specific features while sharing region-invariant features, allowing for training effectively with a more extensive and diverse dataset.
  \item Our in-house data experiments demonstrate improvements with this new framework.
\end{itemize}

\section{Related Work}
\begin{figure}[thb]
\centering
\includegraphics[width=\columnwidth]{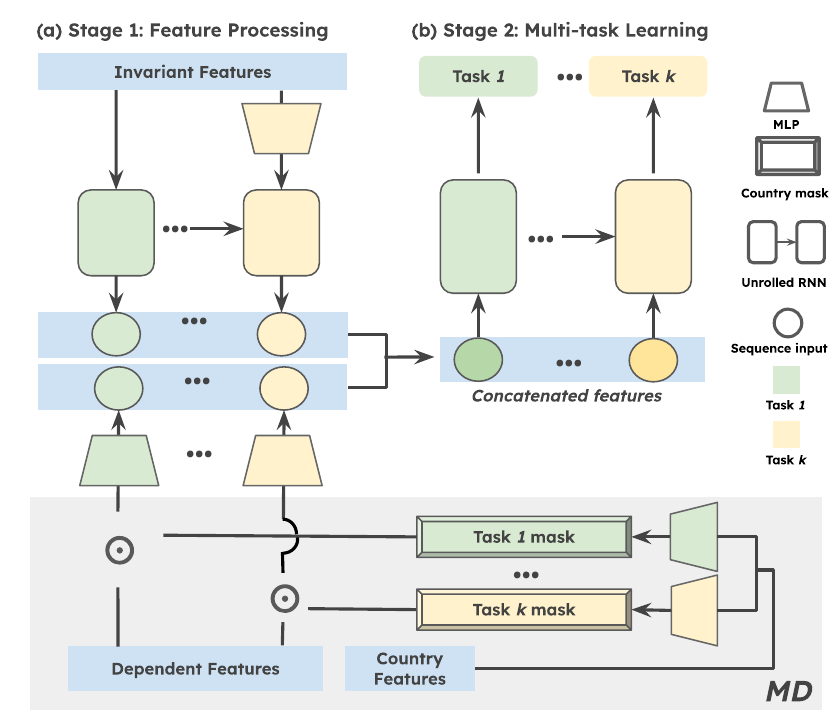}
\caption{\ourmethodspace overall architecture. 
(a) Feature processing. The input is split into three parts: \textit{country features}, \textit{dependent features}, and \textit{invariant features}.\textit{Invariant features} are processed into a sequence input with MLP blocks, and then the features are output from stage 1 RNN as a sequence.
\textit{Country features} and \textit{dependent features} are processed through our multi-distribution (MD) learning module, with each task having its own country mask weights. More details about the multi-distribution adaptor module can be found in Fig.~\ref{fig4_adapt}.
(b) Multi-task Learning. The concatenated features pass through the following RNN layers, providing the model's final output scores for each task. Note that the RNN blocks illustrate the model's architecture, and the number of layers can vary. }
\label{fig3_overall}
\end{figure}

\noindent\textbf{\textit{Multi-task learning (MTL)}} trains models on multiple tasks simultaneously. By sharing information across tasks, the model can learn more robust features, leading to improved performance on each individual task. MTL can be categorized into two types: \textbf{hard parameter sharing} and \textbf{soft parameter sharing}. Hard parameter sharing involves an architecture where certain layers are shared among all tasks in the base model, while other layers remain specific to individual tasks in separate task "towers." The "Shared-bottom" approach~\cite{shared_bottom_caruana1997multitask} is one of the most popular methods within this category. Soft parameter sharing uses trainable parameters to combine each layer’s output with linear combinations. This approach often incorporates the concepts of \textit{experts} and \textit{gates}, which are multi-layer perceptrons (MLPs) in the architecture design. \textit{Experts} are responsible for learning with specific attentions from the features, while \textit{gates} determine how to combine these attentions. Various methods differ based on whether the \textit{experts} and \textit{gates} are shared among tasks or specific to individual tasks, as shown in Fig.~\ref{fig2_mtl_compare}-(a). \Eg, MMoE~\cite{mlmmoe_ma2018modeling} shares all \textit{experts} and \textit{gates} parameters among the tasks; PLE~\cite{ple_tang2020progressive} includes both task-specific and shared \textit{experts} and \textit{gates}; Adatt-sp~\cite{li2023adatt} has task-specific experts, but all gates are shared among tasks. Soft parameter sharing heavily relies on experts and gates for knowledge sharing between multiple tasks. However, many related works often overlook the potential to utilize relationships between tasks in MTL. For tasks with a sequential order, Recurrent Neural Networks (RNNs) offer another method to promote knowledge sharing, which has been less explored.

\noindent\textbf{\textit{Sequence learning in e-commerce}} 
learns user patterns has been explored. For instance, DPN~\cite{zhang2024deep} retrieves target-related user behavior patterns using a target-aware attention mechanism, where user behaviors are represented by their shopping history—a sequence of purchased listings. Similarly, Hidasi \etal~\cite{hidasi2018recurrent} demonstrate the impressive performance of RNNs over classical methods in session-based recommendations. GRU4Rec~\cite{hidasi2015session} takes the listing from the current event in the session and outputs a set of scores indicating the likelihood of each listing being the next in the session. However, these related work primarily focus on learning from listing interaction histories. To the best of our knowledge, our work is the first to treat tasks themselves as a sequence in the context of e-commerce.

\noindent\textbf{\textit{Multi-distribution learning}} trains models using data from various sources, each with distinct feature distributions. Multi-regional data is a typical multi-distribution input, where prior work has primarily focused on the language-agnostic aspect, aiming to learn a unified embedding space without language bias~\cite{ahuja2020language}. In contrast, our approach results diversification by incorporating regionally distinct signals. \textbf{Domain generalization (DG)} and \textbf{transfer learning} are the two most relevant topics for dealing with data from various distributions. Domain Generalization (DG) refers to the ability of a model to generalize well to new, unseen domains without access to data from those domains during training~\cite{zhou2022domain}. DG aims to develop models that perform robustly across various domains by leveraging only the data from the source domains. DG has shown promising results, particularly in the computer vision field, where training images may come from different types (e.g., photos, cartoons, sketches)~\cite{li2017deeper,li2018learning}. Several new DG methods have recently emerged and shown strong performance. For example, SWAD~\cite{cha2021swad} enhances generalization by performing stochastic weight averaging on model weights during training, helping to find flat loss minima. MIRO~\cite{cha2022domain} leverages pre-trained models as constraints to guide the training of the target model, thereby learning more robust and generalizable representations. However, our problem setting differs from DG. While DG focuses on learning domain-invariant representations, our approach to training with multi-regional source data aims to retain both domain-invariant and domain-dependent features. In other words, the model's predictions are expected to vary for different regions, unlike DG, which aims for a unified output. Transfer Learning involves leveraging knowledge from a pre-trained model on one task or domain (the source) to improve performance on a different but related task or domain (the target). Adapting models for each region, especially for unbalanced regional data where very limited data is available, is less optimal. Moreover, adapting each single model is time and energy-intensive in real-world scenarios~\cite{vzliobaite2015towards}. Therefore, developing a module to effectively learn from multi-regional source data is both under-explored and of vital importance.

\section{Method}
In this section, we introduce our \ourmethodspace framework, which includes two model components: a multi-task learning architecture \textbf{SEQ} and a multi-distribution learning module \textbf{MD}. We provide formal definitions for the problem followed by detailed explanations for our framework in the subsections.

\begin{figure*}[thb]
\centering
\includegraphics[width=0.8\textwidth]{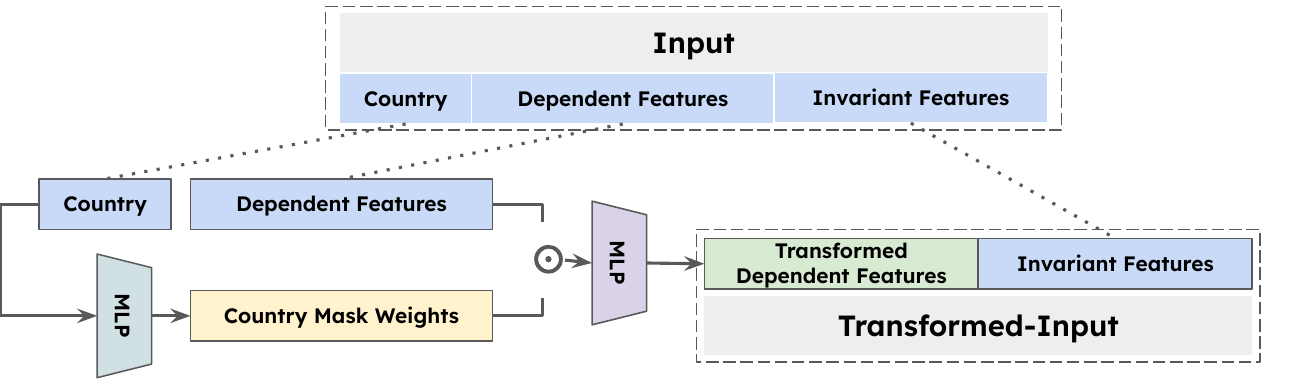}
\caption{Multi-Distribution Adaptor Module (MD). The input is broken down into three parts: \textit{Country features} (the cause of the distribution difference), \textit{dependent features} (the features with multi-distributions), and \textit{invariant features} (the features with consistent distributions). The \textit{country features} generate a weight mask through an MLP block, which is then element-wise multiplied with the \textit{dependent features}. This product feature is processed through an MLP, producing \textit{transformed dependent features} that are assumed to be invariant. These are then concatenated with the original \textit{invariant features} from the input to create the transformed input. This transformed input can then be passed to any MTL models for further processing.}
\label{fig4_adapt}
\end{figure*}

\subsection{Problem Definition}
Consider an online shopping dataset that records the journey of users querying an listing and interacting (\eg, click, purchase) with the returned listings. 
Let $ D = \{(X_i, Y_i)\}_{i=1}^n $ be the dataset with $ n $ samples, where 
$ X = (x_u^m, x_l^p) $, $x_u^m$ refers to the $m$-dimensional features about the \textit{user} and \textit{query}, $x_l^p$ refers to the $p$-dimensional features about the target \textit{listing}, and
$ Y = \{y_i\}_{i=1}^k $ is the score set for $k$ tasks. 
The score for each task is calculated based on the user interaction sequences. A complete sequence would be ["click", "add to cart", "purchase"]. The last action in this sequence represents the final step. For example, if the sequence is ["click", "add to cart"], it means the user clicked on the listing and added it to the cart but did not purchase it. If none of these actions occurred, the sequence is ["no interaction"]. We assign specific scores to each action ("no interaction", "click", "add to cart", "purchase"), and the final task score is a combination of these action scores.

The multi-task learning architecture SEQ focuses on making predictions for the $k$ tasks simultaneously given a single input $X$. Meanwhile, the multi-distribution learning module MD is designed for unified learning across the entire input set $\{(X_i)\}_{i=1}^n$, where the distribution of $X$ for certain regions shows significant differences compared to other regions. (See Fig. \ref{fig1_motivation}-(b) for examples.) The multi-task learning architecture and multi-distribution learning module can be applied separately. We combine these two parts in our final framework and Fig.~\ref{fig3_overall} shows the overall structure. 



\subsection{Learning Multi-Task as A SEQuence}
\label{method_seq}
Some tasks naturally form a sequence, e.g., \textit{click}, \textit{add to cart}, \textit{purchase}, where each action occurs in a sequential order, conditional on the previous ones. However, most multi-task learning architectures do not account for the sequential nature of the problem, making the output tasks order-agnostic and interchangeable. 


Introducing "order" into multi-task learning offers several benefits. First, sequential ordering allows the model to \textbf{prioritize more complex tasks} later in the sequence. In e-commerce, those later tasks (\eg \textit{purchase}) are often more critical than earlier (\eg \textit{click}) task because of their higher monetization values. At the same time, the data sparsity of the \textit{purchase} task makes it more difficult to optimize. 
By establishing a sequence, knowledge from earlier (and typically easier) tasks can be used to address later (and often harder) tasks. Second, sequential ordering \textbf{facilitates the transfer or addition of new tasks}. Since the model learns tasks in a "continuous" manner, adding new tasks in the sequence requires minimal training cost.
Journey Ranker~\cite{tan2023optimizing} recognized the importance of task order having each task model predict the conditional probability based on the previous task. However, the MLP components in their model are isolated, not fulling utilizing the knowledge exchange of the sequential tasks.

To address this, we connect RNNs~\cite{rnn_cho2014learning} with multi-sequential-task learning. In RNN~\cite{rnn_cho2014learning}, the prediction of later tokens is based on previous tokens; similarly, our predictions for later user actions are conditioned on previous actions. In RNN~\cite{rnn_cho2014learning}, each layer shares the same set of weights, with the only difference being the input token and the hidden input from previous tokens. In our approach, as shown in Eq.~\ref{eq:seq_process}, we process the single input feature through an MLP for each token, transforming the input feature specifically for each task (see Fig. \ref{fig3_overall}-(a)). The hidden input can be seen as the knowledge passed down from previous actions. Gated Recurrent Unit (GRU)~\cite{gru_chung2014empirical} is applied in our SEQ architecture.

\begin{equation}
[X_i^0, ..., X_i^{k-1}] = [X_i, ..., MLP^{k-1}(X_i)]
\label{eq:seq_process}
\end{equation}

Fig. \ref{fig3_overall} shows our sequential task learning together with MD module. Given a single input feature, the first step is passing it through $k-1$ MLPs to create a length-$k$ sequence, where $k$ is the number of tasks. After passing through multiple layers of RNN, the output scores are in sequence form, with each score token corresponding to a task.

To further strengthen the learning with sequence, we add the \textbf{Descending Probability Regularizer}~\cite{tan2023optimizing}. Based on the prior knowledge that the probability of a sequence of actions decreases from the beginning to the end (i.e., the probability of a user "clicking" the listing is greater than or equal to the probability of "purchasing"), we add a sigmoid multiplication at the end of the output. Each output score is activated with a sigmoid function and then multiplied by the previous sigmoid scores. As shown in Eq.~\ref{eq:descend_reg}, the score for task $m$, $\tilde{y_m}$ is the product of the sigmoid activations of the logits $l$ from all previous tasks. This ensures that the output probabilities of later actions are always smaller than those of previous actions, aligning with the prior knowledge.

\begin{equation}
\tilde{y_m}= \prod_{i=1}^{i=m} sigmoid(l_{i})
\label{eq:descend_reg}
\end{equation}

\subsection{Learning with Multi-Distribution Input}
Looking at the distribution of each raw input feature, we noticed that there are multi-distributions for certain features (\eg average number of purchases, see examples in Fig.\ref{fig1_motivation}-(b)). If the goal of training a machine learning model is to learn the transition from a input distribution to the output distribution, then this multi-distribution will pose significant challenges to the model, ultimately leading to a failure in learning~\cite{peng2024sample}.

Fig.~\ref{fig4_adapt} shows the overall structure of the multi-distribution adaptor module. We first break the input features into three parts: \textit{country features}  (which is the deciding factor of the distribution shift), \textit{dependent features} (with distribution shifts across countries), and \textit{invariant features} (which are country-agnostic features). The feature split is done in a heuristic way: country features are manually selected, and the \textit{dependent features} and \textit{invariant features} are separated with a distribution distance threshold. \textit{i.e.}, when the average of the distribution distance among all countries is greater than a certain threshold, the feature is categorized as a \textit{dependent feature}.

After splitting the input features, different operations are applied to these three groups of features. \textit{Country} features are used to generate \textit{country mask weights} for the \textit{dependent features}. \textit{Country mask weights} have the same dimension as the \textit{dependent features}, and elementwise-multiplication is performed between the mask and \textit{dependent features}. The multiplied input is fed into an MLP, which transforms the output into invariant features. These are then concatenated with the \textit{invariant features} from the original input, resulting in a transformed input with consistent distributions.

This multi-distribution adaptor module MD can be "plug-and-play" for all MTL frameworks. Adding this module directly after the input and then sending the transformed input to the model is clean and simple. We also explore other options for combining this adaptor module with our sequential task learning framework, as shown in Fig.~\ref{fig3_overall}. Instead of concatenating the transformed dependent features with the input feature directly, we can concatenate them with the invariant feature model output from the previous layers. Block (b) in Fig.~\ref{fig3_overall} shows how the multi-distribution module works in our sequential learning architecture. Each task has its own country mask. For a single input (\textit{country features}, \textit{dependent features}) transformed with $k$-task country masks, the output is also a length-$k$ input sequence. Concatenated with the invariant feature output, the new input features can be processed with the following sequential learning layers to finally get the task scores.
\section{Experiments}
\begin{table*}[tbh]
    \centering
    \setlength{\tabcolsep}{5.5pt}
    \caption{Multi-task learning performance. Results are reported with respect to the shared-bottom model baseline, with the best results marked in bold. State-of-the-art methods are listed in (a) and our models in (b). Our SEQ model outperforms all baselines in (a) across all tasks and platforms. Adding the multi-distribution learning module \textit{MD} further enhance the performance. See Section ~\ref{subsec:result-mtmd} for further discussion.}  
    \begin{tabular}{rlcccc}
    \hline
    & & \multicolumn{2}{c}{Click Task} & \multicolumn{2}{c}{Purchase Task}\\
    & Platform & Web  & App    & Web  & App\\
    \hline
        \textbf{(a)} & MLMMoE~\cite{mlmmoe_ma2018modeling}   & -0.043\% &-0.315\% &+1.027\% & +0.553\%\\
                     & PLE~\cite{ple_tang2020progressive}  & -0.450\%& -0.298\% & +0.790\% & +0.512\% \\
                     & AdaTT~\cite{li2023adatt}   & -0.001\%& -0.541\%& +0.572\%& +0.556\%\\
        \hline
        \textbf{(b)} & SEQ & \textbf{+0.618\%} & \textbf{+0.476\%}& +1.305\% & +1.426\% \\
        & SEQ+MD  & +0.170\% & +0.091\% & \textbf{+1.705\%} & \textbf{+1.952\%} \\
        \hline
    \end{tabular}
    \label{tab:mtl_results}
\end{table*}

\begin{table*}[tbh]
    \centering
    \setlength{\tabcolsep}{5.5pt}
    \caption{Multi-distribution learning module \textit{MD} performance. Results are reported based on the improvements over shared-bottom model~\cite{shared_bottom_caruana1997multitask} baseline, with the best results marked in bold. Applying the \textit{MD} module to state-of-the-art MTL methods demonstrates varying degrees of overall improvement (Percentage changes with regard to no-MD baselines are marked in \textcolor{teal}{green} for improvements and \textcolor{red}{red} for declines in performance). See Section ~\ref{subsec:result-md} for further discussion.} 
    \begin{tabular}{rcccc}
    \hline
    & \multicolumn{2}{c}{Click Task} & \multicolumn{2}{c}{Purchase Task}\\
    Platform & Web  & App   & Web  & App\\
    \hline
    MLMMoE~\cite{mlmmoe_ma2018modeling}   & -0.043\% &-0.315\% &+1.027\% & +0.553\%\\
                     MLMMoE~\cite{mlmmoe_ma2018modeling} + MD  & \textbf{+0.629}\% $_{\textcolor{teal}{0.673\%}}$&  \textbf{+0.291\%} $_{\textcolor{teal}{0.607\%}}$& +0.129\% $_{\textcolor{red}{0.889\%}}$&  -0.025\% $_{\textcolor{red}{0.575\%}}$\\
                     PLE~\cite{ple_tang2020progressive}  & -0.450\%& -0.298\% & +0.790\% & +0.512\% \\
                     PLE~\cite{ple_tang2020progressive} + MD & -0.023\% $_{\textcolor{teal}{0.429\%}}$ & -0.113\% $_{\textcolor{teal}{0.186\%}}$ & \textbf{+1.958 \%}$_{\textcolor{teal}{1.159\%}}$&  \textbf{+1.891\%}$_{\textcolor{teal}{1.372\%}}$\\
                     AdaTT~\cite{li2023adatt}   & -0.001\%& -0.541\%& +0.572\% & +0.556\%\\
                     AdaTT~\cite{li2023adatt} + MD  & +0.477\%$_{\textcolor{teal}{0.478\%}}$ & -0.115\%$_{\textcolor{teal}{0.428\%}}$& +1.060\%$_{\textcolor{teal}{0.486\%}}$&  +0.863\%$_{\textcolor{teal}{0.306\%}}$ \\
        \hline
    \end{tabular} 
    \label{tab:cweight_results}
\end{table*}

To evaluate our methods, we conducted experiments on our offline in-house datasets. Four baseline methods were selected for comparison. The Shared-Bottom model~\cite{shared_bottom_caruana1997multitask} is used as the baseline for all other models, as it represents the most fundamental architecture in multi-task learning (MTL). Results are reported as \textbf{changes relative to the Shared-Bottom model, with its performance marked as the 0\% reference point}. The other methods implemented for reference are MLMMOE~\cite{mlmmoe_ma2018modeling}, PLE~\cite{ple_tang2020progressive}, and Adatt~\cite{li2023adatt}. Details of the baselines are described in Sec.~\ref{subsec:baselines}.


We used 14 days of offline in-house data for training and three days of data for evaluation, and we report the relative increase in the average Normalized Discounted Cumulative Gains ($NDCG$)~\cite{ndcg_valizadegan2009learning} in the result tables (see Sec.~\ref{subsec:metrics} for more details). Due to the varying nature of different traffic sources, the results are divided into two sections: Webpage search traffic (Web), and Mobile App search traffic (App). We track multi-tasks across all traffics.

The results focus on two main areas: the effectiveness of the sequential learning architecture for MTL and the "plug-and-play" multi-distribution learning module for SOTA MTL methods. Ablation studies and alternative designs are discussed in Sec.~\ref{sec:discussions}.

\subsection{Baseline Models}
\label{subsec:baselines}
We select a few state-of-the-art multi-task learning methods without any multi-distribution adjustments as the baselines. For multi-distribution learning challenge, most related work~\cite{cha2022domain, cha2021swad} focuses on learning invariant features, whereas our goal is to better capture regional preferences. Thus, we use training with single or multi-distribution data as the baselines for multi-distribution learning comparisons. 

\noindent\textbf{Shared-bottom}~\cite{shared_bottom_caruana1997multitask} is a hard parameter sharing method in MTL. It consists of a shared bottom layer for all tasks, followed by separate "tower" layers for each task, which extend from the shared-bottom output. Both the "bottom" and the "towers" are MLPs, with no knowledge sharing beyond the shared-bottom.

\noindent\textbf{MLMMOE}~\cite{mlmmoe_ma2018modeling} is a soft parameter sharing method in MTL. It features \textit{experts} and \textit{gates}, which are MLPs within the architecture. "ML" refers to multiple layers; except for the top task-specific gates, all other \textit{experts} and \textit{gates} are shared among tasks.

\noindent\textbf{PLE}~\cite{ple_tang2020progressive} is another soft parameter sharing method in MTL. It includes two types of \textit{experts} and \textit{gates}: task-specific and task-shared. Task-specific \textit{experts} learn only for their individual tasks, and task-specific \textit{gates} accept input exclusively from the same task \textit{expert} or the shared \textit{expert}.

\noindent\textbf{Adatt-sp}~\cite{li2023adatt} is also a soft parameter sharing method in MTL. All \textit{experts} are task-specific, while all \textit{gates} take outputs from all \textit{experts} as their input.

\subsection{Datasets and Metrics}
\label{subsec:metrics}
We exclusively use our in-house data for experiments because public search datasets~\cite{li2020improving} often omit feature details for data security reasons. This omission makes it difficult to isolate country features and generate accurate country mask weights. Our offline in-house dataset contains over 20 million <user, query, listing> interaction sequences from 10 regions and 2 platforms. Unless otherwise specified, we train the models with data from all regions and platforms. Results are evaluated separately for each platform. Normalized Discounted Cumulative Gain ($NDCG$)~\cite{ndcg_valizadegan2009learning} is our evaluation metric, commonly used for measuring the effectiveness of search engines by summing the gain of the results, discounted by their ranked positions. The rankings of the search listings are ordered by the output scores from the model, and $NDCG$ is calculated based on the user interaction sequences. As discussed in Sec.~\ref{method_seq}, e-commerce prioritizes the \textit{purchase} task over \textit{click}, making \textit{purchase-ndcg} our prioritized metric for model evaluation.

\subsection{Results}
\textbf{SEQ.}
\label{subsec:result-mtmd}
Table~\ref{tab:mtl_results} presents the multi-task learning performance on \textit{click} and \textit{purchase} tasks across different platforms. State-of-the-art MTL baseline methods demonstrate various improvements in the \textit{purchase} task but show a slight decline in the \textit{click} task. In contrast, our SEQ model shows improvement across all tasks, adding MD module (SEQ+MD) achieves the best $NDCG$ on the critical \textit{purchase} task.
We observed a performance drop in the \textit{click} task after adding the MD module to SEQ, making the final \textit{click} performance only slightly positive compared to the share-bottom baseline. This may be due to the \textit{click} data being noisier and having higher variance. Another possible explanation is that the region-dependent features isolated by the MD module are more closely related to user/listing purchase history, which may have a greater impact on the purchase task.

\noindent\textbf{MD: Multi-Distribution Learning Module}.
\label{subsec:result-md}
Table~\ref{tab:cweight_results} illustrates the effectiveness of our multi-distribution learning module as a "plug \& play" component for state-of-the-art MTL methods. The adapted models demonstrate overall improvements, with PLE~\cite{ple_tang2020progressive}+MD achieving the best performance for the \textit{purchase} task across all platforms. These results validate that our MD module can significantly enhance MTL performance.

\section{Discussions}
\label{sec:discussions}

\subsection{Will the sequential learning model benefit from more tasks?}
\label{subsec:3t}

\begin{table*}[h]
    \centering
    \setlength{\tabcolsep}{5.5pt}
    \caption{Three-task learning performance. Results are reported based on the shared-bottom~\cite{shared_bottom_caruana1997multitask} model baseline, with the best results marked in bold. Upon adding an additional task, \textit{add to cart}, our SEQ+MD model continues to outperform others, demonstrating even larger performance gains compared to the two-task learning scenario. See Sec.\ref{subsec:3t} for the discussion.}  
    \begin{tabular}{lcccccc}
    \hline
    & \multicolumn{2}{c}{Click Task} & \multicolumn{2}{c}{Add to Cart Task} & \multicolumn{2}{c}{Purchase Task}\\
    Platform & Web  & App  & Web  & App & Web  & App\\
    \hline
    MLMMoE~\cite{mlmmoe_ma2018modeling}   & -0.025\% & -0.627\% & +0.885\% & +0.883\% & +0.596\% & +0.769\%\\
    PLE~\cite{ple_tang2020progressive}   & -0.728\% & -0.458\% & +0.672 \% & +0.475\% & +1.247 \% & +1.396\%\\
    AdaTT~\cite{li2023adatt}   & \textbf{+0.163\%} & -0.054\% & +0.459\% & +0.698\% & +0.901\% & +1.265\%\\
    SEQ+MD & -0.955\% & \textbf{+0.104\%} & \textbf{+0.990\%} & \textbf{+1.029\%} & \textbf{+1.731\%} & \textbf{+2.342\%} \\
    \hline
    \end{tabular}
    \label{tab:3t}
\end{table*}

A significant advantage of learning multi-task sequences lies in the inherent properties of RNNs, where weights are shared across all tokens in the sequence. This has two main benefits. First, it reduces redundant calculations among related tasks. For instance, tasks like \textit{click} and \textit{purchase} share many commonalities in the buyer's decision process, \ie a listing that a user clicks on is also likely to be purchased. Second, by reinforcing the connections between tasks, later tasks in the sequence can be learned more effectively by decomposing them and beginning with easier tasks. As the sequence progresses, task difficulty can be seen as increasing, with earlier tasks acting as processors for the later ones. This recurrent learning process, from easier to harder tasks, is advantageous. For example, predicting which listing is likely to be \textit{purchased} is challenging, but if the model starts by learning \textit{click} behavior, it can learn better. We hypothesize that the sequential learning model will benefit from more tasks. In our experiment, we add an \textit{add to cart} task between the \textit{click} and \textit{purchase} sequence to better reflect the buyer's shopping journey. The results in Table~\ref{tab:3t} support this hypothesis.

\subsection{Transferability from two-task to three-task}
\label{subsec:2t_3t}
\begin{figure}[h]
\centering
\includegraphics[width=\columnwidth]{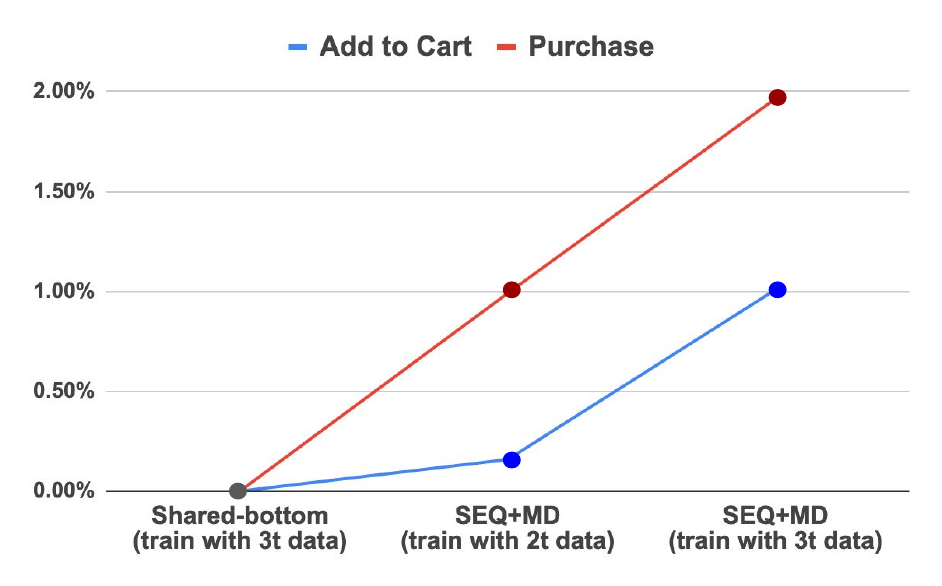}
\caption{Transferability of SEQ+MD from two-task to three-task models is evaluated by comparing the performance of shared-bottom~\cite{shared_bottom_caruana1997multitask} and SEQ+MD models trained on three-task data with the SEQ+MD model trained on two-task data. Remarkably, despite the SEQ+MD model not being trained on \textit{add to cart} data, it still shows improved performance on the \textit{add to cart} and \textit{purchase} tasks when compared to the shared-bottom~\cite{shared_bottom_caruana1997multitask} model. See Sec.~\ref{subsec:2t_3t} for the discussion.
}
\label{fig_2t3t}
\end{figure}

An important consideration for multi-task models is how easily they can adapt to additional tasks. In terms of transferability across different numbers of tasks, the SEQ+MD model demonstrates a significant advantage. Adding new tasks requires almost no increase in parameters compared to the state-of-the-art models which increase parameter size by 30\% on average. Moreover, reusing weights trained on previous tasks can also lead to improved performance in new task evaluations. Figure~\ref{fig_2t3t} illustrates the performance comparison of evaluating a three-task setup using weights from a two-task model. Remarkably, even without having seen the \textit{click} data during training, the model's performance on both \textit{click} and \textit{purchase} tasks still surpasses that of the baseline model trained on three tasks.

\subsection{Ablation studies}
\label{subsec:regularizer}
\begin{figure}[h]
\centering
\includegraphics[width=\columnwidth]{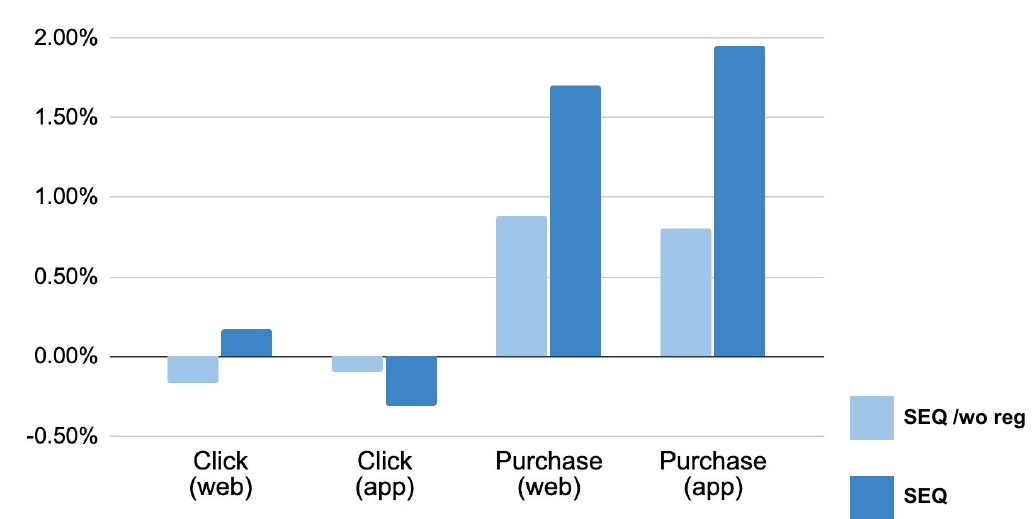}
\caption{The impact of adding the \textit{descending probability regularizer} in the SEQ model. Results are reported based on the improvements over shared-bottom model~\cite{shared_bottom_caruana1997multitask} baseline. Light blue represents the SEQ model without the regularizer, while dark blue indicates the model with the regularizer. The regularizer enhances performance, with noticeable improvements in the purchase task. See Sec.~\ref{subsec:regularizer} for the discussion.
}
\label{fig_reg}
\end{figure}

Learning multi-task as a sequence not only enhances knowledge sharing among tasks but also simplifies the integration of output regularization. In our SEQ design, we incorporate a \textit{descending probability regularizer} that enforces the model to output task scores in a non-increasing order. This regularization is based on the observation that the probability of a user purchasing a listing cannot exceed the probability of them clicking on it, as a click typically precedes a purchase. The results in Fig.~\ref{fig_reg} demonstrate the effectiveness of this regularizer. 

\begin{figure*}[h]
\centering
\includegraphics[width=0.8\textwidth]{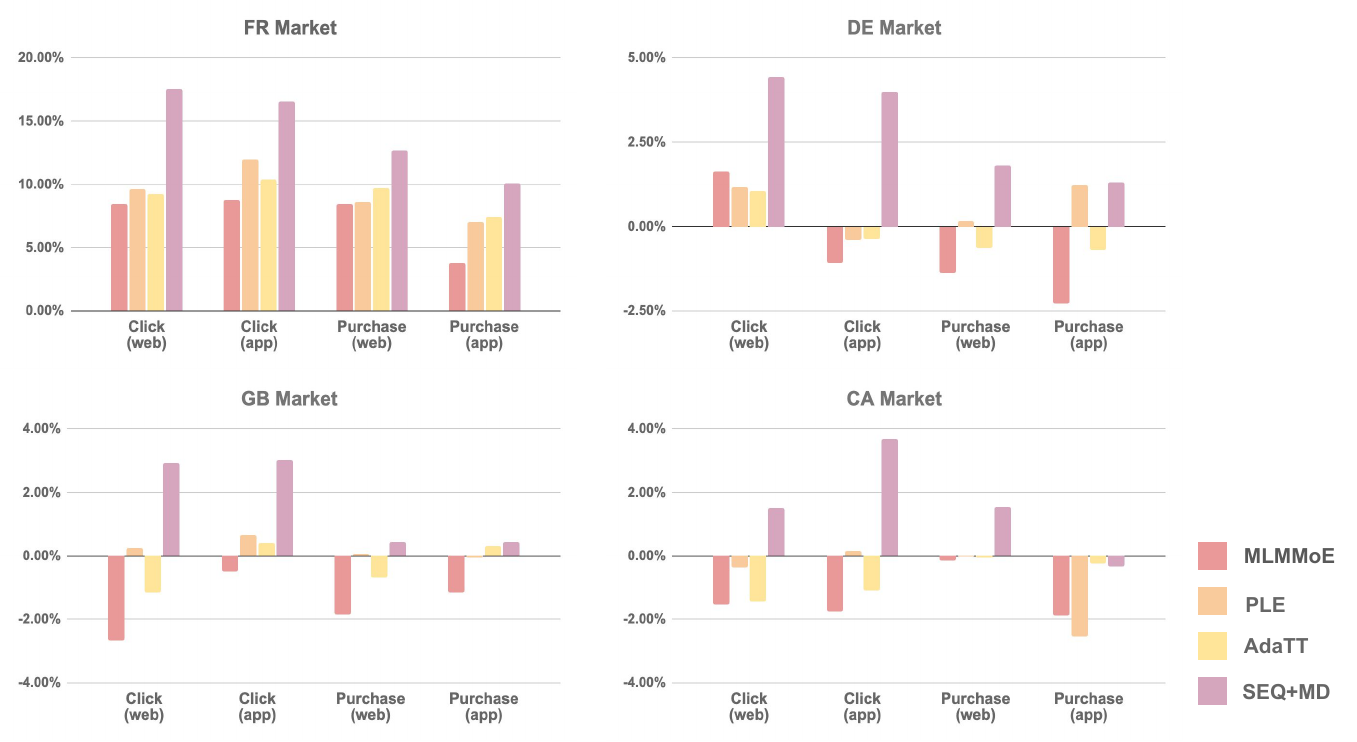}
\caption{Comparisons with training on single regional data. We compare our method SEQ+MD training on all regional data with state-of-the-art MTL models training on the single regional data. Results are reported based on the shared-bottom~\cite{shared_bottom_caruana1997multitask} model baseline training with single regional data. The evaluation in each single region shows the imbalance performance of the baseline methods and our model SEQ+MD constantly performs better among all four regions. See Sec.~\ref{subsec:regional} for the discussion.}
\label{fig_region_barchart}
\end{figure*} 

\subsection{How effective is the MD module when compared to models trained with single regional data?}
\label{subsec:regional}

\begin{figure}[htb]
\centering
\includegraphics[width=0.8\columnwidth]{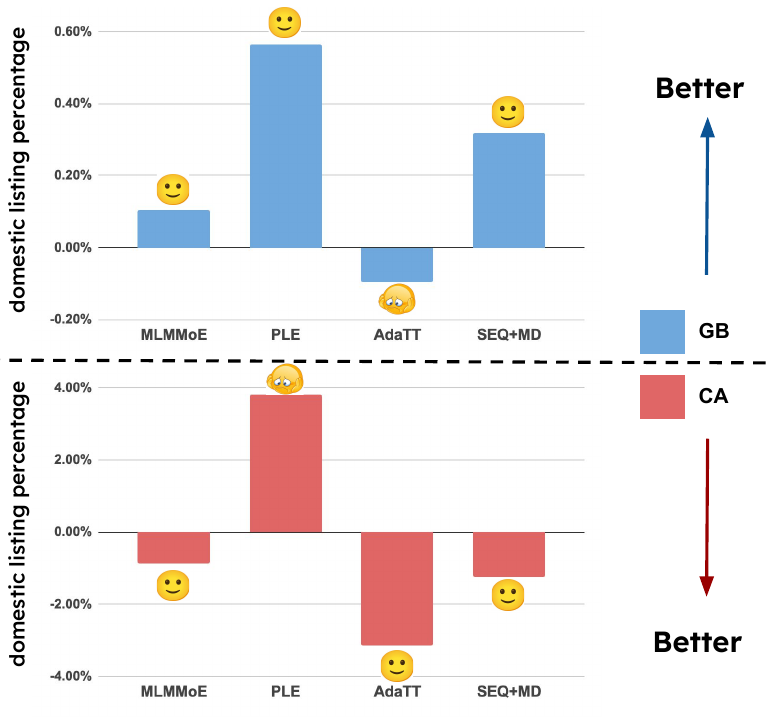}
\caption{Domestic listing percentage changes compared to the shared-bottom model~\cite{shared_bottom_caruana1997multitask} baseline are illustrated for two representative regions: CA and GB. CA buyers tend to favor international listings, while GB buyers prefer domestic options. PLE~\cite{ple_tang2020progressive} and AdaTT~\cite{li2023adatt} show minimal regional differentiation, with AdaTT~\cite{li2023adatt} consistently returning more domestic listings and PLE~\cite{ple_tang2020progressive} returning fewer. In contrast, our SEQ+MD model consistently aligns better with regional preferences, demonstrating superior performance in fitting local market trends. See Sec.~\ref{subsec:regional} for the discussion.
}
\label{fig_domestic}
\end{figure}

Fig.~\ref{fig_region_barchart} compares state-of-the-art baseline models trained on single regional (user region) data with our SEQ+MD model trained on data from all regions. Each evaluation is performed per region. To ensure similar dataset sizes between a single region and all regions, we use different training set lengths for each setting. For our SEQ+MD model, we use 7 days of data from all regions. For regions with larger group of users like GB, we use 14 days of data, while smaller regions like CA and FR require 56 days, and DE uses 42 days.


The results per region highlight the performance imbalance of the baseline models. For example, PLE~\cite{ple_tang2020progressive} shows relatively good performance in DE and FR regions but performs poorly in GB. In contrast, our SEQ+MD model, trained with mixed distributions, shows almost all positive results across all tasks and platforms.

Our SEQ+MD model also demonstrates a superior ability to align with regional preferences compared to other baselines. Figure~\ref{fig_domestic} illustrates the changes in the percentage of domestic listings relative to the shared-bottom~\cite{shared_bottom_caruana1997multitask} model baseline (All models are trained with all-regional data.). Our in-house analysis shows distinct regional preferences in CA and GB, where CA buyers tend to favor international listings, while GB buyers lean towards domestic options. However, Fig.~\ref{fig_domestic} shows that PLE~\cite{ple_tang2020progressive} consistently returns fewer domestic listings, while AdaTT~\cite{li2023adatt} consistently returns more, regardless of these regional preferences. In contrast, our SEQ+MD model effectively captures these regional trends, providing more accurate rankings that better align with the buyers' preferences.

\section{Conclusion}
In this paper, we introduce the SEQ+MD framework, which integrates sequential learning for multi-task problems with multi-distribution data. The two key components, SEQ and MD when combined, can be applied independently but yield better results on the more critical and complex tasks. The motivation behind learning multi-task as a sequence stems from the natural sequential order of tasks. Our experiments and analyses highlight two primary benefits:
First, SEQ reduces redundant computation across tasks and enhances transferability between different task sets, requiring minimal additional parameters while effectively utilizing weights from previous models. Second, by breaking down a complex task into simpler subtasks that serve as processors in the sequence, the model demonstrates improved performance on more challenging tasks.
Additionally, our MD module effectively handles multi-distribution data, and as a "plug-and-play" component, it can also enhance the performance of state-of-the-art multi-task learning models.


\noindent\textbf{Future work.} \textbf{1. Improve robustness against noisy data.} Even though the primary goal of our approach is to improve performance on the complex tasks such as \textit{add to cart} and \textit{purchase}, we see opportunities in making SEQ+MD have a neutral impact on \textit{click} compared to SEQ only. 
One hypothesis is that click data tends to be noisier than other tasks, with a significant amount of "false clicks" present, particularly on mobile platforms. For example, users may accidentally click on a listing due to the touch screen's sensitivity. Learning with task-specific noise within a multi-task learning framework could be a valuable direction for future research.
\textbf{2. Generalize multi-distribution data from region-wise to other scenarios.} While this paper focuses on regional differences as an example of multi-distribution, other multi-distribution exists in e-commerce search data. For instance, different platforms (web, app) may show distinct shopping patterns. Extending our MD module to address these scenarios could be a promising research direction.

{\small
\bibliographystyle{ACM-Reference-Format}
\bibliography{sample-base}
}
\end{document}